\documentclass[letterpaper,12pt]{article}
\usepackage{natbib}
\usepackage{apalike}
\usepackage{epsfig}
\usepackage{setstack}
\usepackage{lscape}
\usepackage{wrapfig}
\usepackage{times}
\usepackage{geometry}
\usepackage{wasysym}
\usepackage[utf8]{inputenc}
\usepackage{enumitem,amssymb}
\usepackage{ragged2e}
\newlist{thematic}{itemize}{8}
\setlist[thematic]{label=$\square$}
\usepackage{pifont}

\begin{document}
\raggedright

\huge
Astro2020 APC White Papers \linebreak

Setting the Stage for the Planet Formation Imager \linebreak
\normalsize

\noindent \textbf{Type of Activity:} \hspace*{60pt} $\XBox$ Ground Based Project \hspace*{10pt} $\square$ Space Based Project \hspace*{20pt}\linebreak
$\square$ Infrastructure Activity \hspace*{31pt} $\XBox$ Technological Development Activity \linebreak
  $\square$  State of the Profession Consideration \hspace*{1pt} $\square$ Other {\underline{~~~~~~~~~~~~~~~~~~~~~~~~~~~~~~~~~~~~~~~~~~~~}} \hspace*{40pt} \linebreak
 \linebreak
  
\textbf{Principal Author:}

Name: John Monnier	
 \linebreak						
Institution:  University of Michigan
 \linebreak
Email: monnier@umich.edu    
 \linebreak
Phone:  734-763-5822
 \linebreak
 
\textbf{Endorsers:} (names, institutions, emails)
  \linebreak

Alicia	Aarnio,	University of North Carolina Greensboro,		anaarnio@uncg.edu		\\
Olivier	Absil,	Université de Liège,		olivier.absil@uliege.be		\\
Almudena	Alonso-Herrero	Centro de Astrobiologia, Spain,		aalonso@cab.inta-csic.es		\\
Narsireddy	Anugu,	University of Exeter,		n.anugu@exeter.ac.uk		\\
Ellyn	Baines,	Naval Research Lab,		ellyn.baines@nrl.navy.mil		\\
Amelia	Bayo,	Universidad de Valparaiso,		amelia.bayo@uv.cl		\\
Jean-Philippe	Berger,	Univ. Grenoble Alpes, IPAG,		Jean-Philippe.Berger@univ-grenoble-alpes.fr		\\
William	Danchi,	NASA Goddard Space Flight Center,		william.c.danchi@nasa.gov		\\
Nicholas	Elias,	Commonwealth Computer Research,		nick.elias@oamsolutions.com		\\
Mario	Gai,	Istituto Nazionale di AstroFisica,		mario.gai@inaf.it		\\
Poshak	Gandhi,	University of Southampton,		poshak.gandhi@soton.ac.uk		\\
Tyler	Gardner,	University of Michigan,		tgardne@umich.edu		\\
Douglas	Gies,	Georgia State University,		gies@chara.gsu.edu		\\
Jean-François	Gonzalez,	Univ Lyon,		Jean-Francois.Gonzalez@ens-lyon.fr		\\
Chris	Haniff,	University of Cambridge, UK,		cah@mrao.cam.ac.uk		\\
Sebastian	Hoenig,	University of Southampton, UK,		s.hoenig@soton.ac.uk		\\
Michael	Ireland,	Australian National University,		michael.ireland@anu.edu.au		\\
Andrea	Isella,	Rice University,		isella@rice.edu		\\
Stephen	Kane,	University of California, Riverside,		skane@ucr.edu		\\
Florian	Kirchschlager,	University College London,		f.kirchschlager@ucl.ac.uk		\\
Makoto	Kishimoto,	Kyoto Sangyo University, Japan,		mak@cc.kyoto-su.ac.jp		\\
Lucia	Klarmann,	Max-Planck-Institut für Astronomie,		klarmann@mpia.de		\\
Jacques	Kluska,	KU Leuven,		jacques.kluska@kuleuven.be		\\
Stefan	Kraus,	University of Exeter,		s.kraus@exeter.ac.uk		\\
Lucas	Labadie,	University of Cologne,		labadie@ph1.uni-koeln.de		\\
Jean-Baptiste	Le Bouquin,	University of Michigan,		lebouquj@umich.edu		\\
David	Leisawitz,	NASA Goddard Space Flight Center,		david.t.leisawitz@nasa.gov		\\
Hendrik	Linz,	MPIA Heidelberg,		linz@mpia.de		\\
Bertrand	Mennesson, JPL, 	bertrand.mennesson@jpl.nasa.gov \\
Andreas	Morlok,	Institut für Planetologie Münster,		morlok70@mac.com		\\
Ryan	Norris,	Georgia State University,		norris@astro.gsu.edu		\\
Benjamin	Pope,	NYU,		benjamin.pope@nyu.edu		\\
Luis Henry	Quiroga-Nuñez,	Leiden Observatory,		quiroganunez@strw.leidenuniv.nl		\\
Gioia	Rau,	NASA/GSFC,		gioia.rau@nasa.gov		\\
Zsolt	Regaly,	Konkoly Observatory,		regaly@konkoly.hu		\\
Mark	Reynolds,	University of Michigan,		markrey@umich.edu		\\
Alberto	Riva,	INAF - Osservatorio Astrofisico di Torino,		alberto.riva@inaf.it		\\
Rachael	Roettenbacher,	Yale University,		rachael.roettenbacher@yale.edu		\\
Gail	Schaefer,	CHARA Array, Georgia State University,		schaefer@chara-array.org		\\
Benjamin	Setterholm,	University of Michigan,		bensett@umich.edu		\\
Michael	Smith,	University of Kent,		m.d.smith@kent.ac.uk		\\
Robert	Stencel,	Univ. Denver - Astronomy,		robert.stencel@du.edu		\\
Theo	ten Brummelaar,	CHARA - Georgia State University,		theo@chara-array.org		\\
Konrad R. W.	Tristram,	European Southern Observatory,		konrad.tristram@eso.org		\\
Gerard	van Belle,	Lowell Observatory,		gerard@lowell.edu		\\
Gautam	Vasisht,	Jet Propulsion Laboratory,		Gautam.Vasisht@jpl.nasa.gov		\\
Gerd	Weigelt,	MPI for Radio Astronomy,		weigelt@mpifr.de		\\
Markus	Wittkowski,	ESO,		mwittkow@eso.org		\\

\pagebreak
\justifying

\noindent\textbf{Abstract:}

An international group of scientists has begun planning for the Planet Formation Imager (PFI, www.planetformationimager.org), a next-generation infrared interferometer array with the primary goal of imaging the active phases of planet formation in nearby star forming regions and taking planetary system ``snapshots'' of young systems to understand exoplanet architectures.  PFI will be sensitive to warm dust emission using mid-infrared capabilities made possible by precise fringe tracking in the near-infrared.  An L/M band beam combiner will be especially sensitive to thermal emission from young exoplanets (and their circumplanetary disks) with a high spectral resolution mode to probe the kinematics of CO and H$_2$O gas.  In this brief White Paper, we summarize the main science goals of PFI, define a baseline PFI architecture that can achieve those goals, and identify key technical challenges that must be overcome before the dreams of PFI can be realized within the typical cost envelope of a major observatory.  We also suggest activities over the next decade at the flagship US facilities (CHARA, NPOI, MROI)  that will help make the Planet Formation Imager facility a reality.  {\bf The key takeaway is that infrared interferometry will require new experimental telescope designs that can scale to 8\,m-class with the potential to reduce per area costs by $\times$10, a breakthrough that would also drive major advances across astronomy.}

\vspace{.25in}

\noindent {\bf Key Science Goals and Objectives:}

\begin{wrapfigure}[18]{r}{0.5\textwidth}
\vspace{-.4in}
  \includegraphics[width=.5\textwidth]{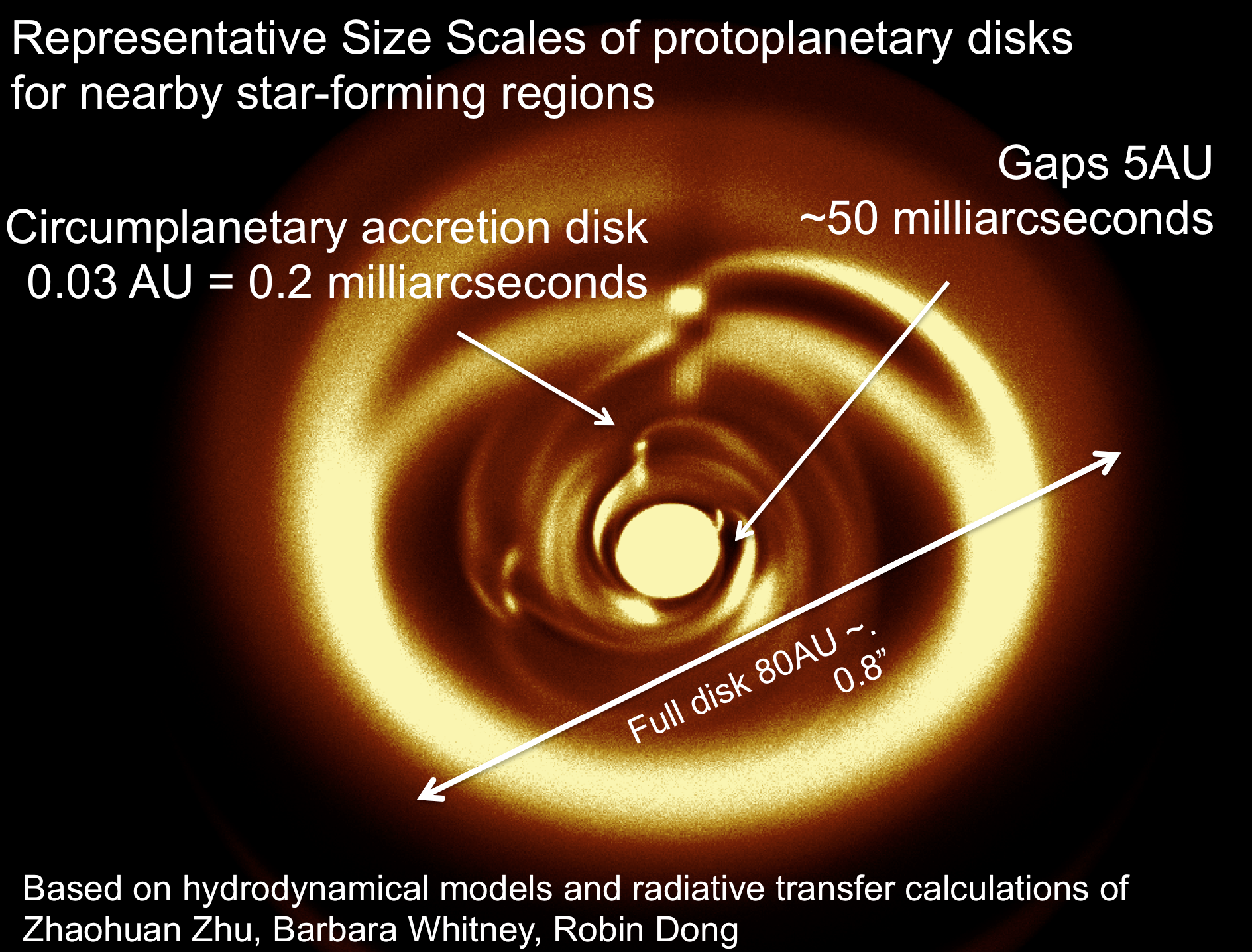}
\caption{\footnotesize Radiative transfer model for an example planet-forming disk \citep{pfimonnier2014,dong2015} with the relevant size scales marked. The primary science driver of the Planet Formation Imager (PFI) is to image scales as large as the whole circumstellar accretion disk down to the circumplanetary accretion disks of individual giant planets.  }
\label{fig:1}       %
\end{wrapfigure}

The Planet Formation Imager (PFI) Project \citep{pfimonnier2014,pfikraus2014,pfiireland2014} was started in 2013 to radically advance the field of planet formation, to image the sub-au spatial scales sufficient to resolve disk gaps cleared by single planets, to detect accretion streams, and to follow the dust and gas all the way down to scales of individual exoplanet Hill Spheres, where disk material is accreted onto young planets themselves.  We introduce the PFI science goals by looking at a radiative transfer image of a planet-forming disk in the thermal infrared (see Figure~\ref{fig:1}).  The protoplanetary disk is approximately 80\,au across, with gaps and structures on the scale of $\sim$5\,au.  We expect a circumplanetary disk to form on scales of 0.03\,au, matching the Hill Sphere for each accreting protoplanet (for a Jupiter-mass planet on a 5\,au radius orbit).  The mid-IR wavelength range efficiently traces emission from small grains from 0.1-10\,au in the disk, complementing mm-wave/radio observations of the large grains.  {\em In the mid-IR, probing scales of 0.1\,au at the distance of even the nearest star-forming regions is far beyond the capabilities of a single telescope and requires an infrared interferometer with kilometric baselines.}

Based on the science case for PFI (presented in ASTRO2020 Science White Papers and in the literature), we summarize the PFI top-level science requirements in Table~\ref{tlsr}.  There are hundreds of young stellar objects with disks that satisfy these requirements within 200~pc and thousands if we move out to the distance of Orion. 
In the next section we propose a specific facility architecture that can meet these top level science requirements at the cost of a typical major astronomical facility.

For the stars in this sample, PFI will be able to detect massive planets at all stellocentric radii, providing a complete census of the exoplanet population down to a certain mass limit. The objects in this sample cover a wide range of evolutionary stages, from the pre-main-sequence ($\sim0.1$\,Myr), transitional disk ($\sim10$\,Myr), to the debris disk phase ($\sim 100$\,Myr). It is expected that planetary systems undergo dramatic changes during this age range, with planets forming in the outer disk and then migrating inwards and outwards due to planet-disk interaction, and planets being rearranged or ejected due to resonances and planet-planet scattering \citep{raymond2006}. By observing planetary systems at different evolutionary stages, PFI will be able to observe directly how these processes alter the exoplanet populations with time. This will provide much-needed input for constraining exoplanet population synthesis models and unveil the dynamical processes that determine the architecture of exoplanetary systems and that shaped also our own solar system.

With sub-milliarcsecond angular resolution and ground-breaking sensitivity, PFI will also image dust tori around Active Galactic Nuclei, AGB star mass-loss, magnetic spots on main sequence stars, trace stellar orbits around the Galactic Center, measure diameters of young stars themselves, and more.
{\em Additional information on the capabilities of the Planet Formation Imager can be found in a recent paper in {\em Experimental Astronomy} \citep{monnier2018pfi} and in two submitted ASTRO2020 White Papers: ``Imaging the Key Stages of Planet Formation" (lead: Monnier) and ``The Future of Exoplanet Direct Detection" (lead: Monnier).
}

\begin{table}
\begin{center}
\caption{Top-level Science Requirements (Minimum Goals)}
\begin{tabular}{|l|c|c|}
\hline
 Parameter & Dust Imaging & Young Exoplanets\\
\hline
Wavelengths & 5-13\,$\mu$m & 3-5\,$\mu$m \\
Typical Source Distance & 140\,pc & 50-500\,pc \\
Spatial Resolution & 2\,mas $\equiv$ 0.3~AU & 0.7\,mas $\equiv$ 0.1~AU\\
Sensitivity & Integrated $m_N\sim$7 & $m_L\sim18.5$ (Point source @5$\sigma$) \\
Goal Surface Brightness (K) & 150\,K  &  $--$\\
Spectral Resolving Power& R$>100$ & R$>100$ and R$>10^5$ (spectroscopy) \\
Field-of-view &  $>0.15"$  & $>0.15"$ \\
Fringe tracking limit & $m_H>9$ & $m_H>9$ \\
Fringe tracking star & $\phi<0.15\,$mas & $\phi<0.15\,$mas \\
\hline
\end{tabular}
\label{tlsr}
\end{center}
\vspace{-.2in}
\end{table}

\vspace{0.25in}

\noindent {\bf Technical Overview:} 

Members of the PFI Technical Working Group have studied how to best achieve the PFI top-level science requirements (see Table~\ref{tlsr}) and have presented their work in various SPIE papers \citep{pfimonnier2014,pfiireland2014,pfimonnier2016,pfikraus2016,pfiireland2016,pfiminardi2016,pfimozurkewich2016,pfibesser2016,pfipetrov2016,monnier2018pfi, monnier2018spie,ireland2018spie,zunigo2018spie,defrere2018spie,quanz2018spie,pedretti2018spie,besser2018spie,michael2018spie,tristam2018spie}.  Here we give an overview of our current conclusions regarding the optimal PFI telescope array architectures.

\begin{table}
\begin{center}
\caption{Technical Description of Reference PFI Architectures\label{pfiref}}
\begin{tabular}{|l|c|c|}
\hline
Parameter & 12$\times$3\,m PFI & 12$\times$8\,m PFI \\
\hline 
Number of Telescopes & 12  & 12 \\
Telescope Diameter & 3\,m  & 8\,m \\
Maximum Baseline & 1.2\,km & 1.2\,km \\
Goal Science Wavelengths & 3--13\,$\mu$m  &3--13\,$\mu$m \\
Fringe-tracking wavelengths & 1.5--2.4\,$\mu$m & 1.5--2.4\,$\mu$m \\
Fringe tracking limits (point source) & $m_H<$13  & $m_H<$15 (AO-dependent)  \\
Point source Sensitivity (10$^4$\,s integration) & 18.1 (L), 12.2 (N) & 20.2 (L), 14.3 (N)   \\
Surface Brightness Limit (10$^4$\,s, $B=1.2$\,km) & 150\,K (N)& 125\,K (N) \\
Field-of-view & 0.25" (L), 0.7" (N) & 0.09" (L), 0.25" (N)\\
Note & w/ Nulling (2-4\,$\mu$m) &  w/ Nulling (2-4\,$\mu$m) \\
Cost estimate$^*$ & \$250M & \$600M$^{**}$ \\
\hline
\end{tabular}
\label{pfispecs}
\end{center}
{* Interferometry cost model is site-dependent and generally highly uncertain; assumptions were presented originally by \citet{ireland2016spie} and updated by \citet{monnier2018pfi}.\\
{**} This informal cost estimate is actually for 12$\times$6.5m telescopes, assuming roughly x2 cost reduction from mass-production efficiencies of an existing telescope design
\citep{kingsley2018spie}. }
\end{table}

\citet{pfiireland2016} laid out the basic SNR equations for PFI used in our limiting magnitudes and sensitivity calculations. For background-limited observations (i.e., not limited by systematics or nulling performance), a simplified version of the SNR equation for point source detection is:
\begin{equation}
{\rm SNR}_{\rm pt} \propto \frac{\sqrt{N_t} D^2 \sqrt{t}}{\sqrt{B(T)}},
\end{equation}
where $N_t$ is the number of telescopes, $D$ is the telescope diameter, $B(T)$ is the background emissivity (if thermal, $T$ is temperature).  Notice that the noise level is independent of the telescope size in the background limit since a diffraction-limited system has constant \'etendue ($\Delta\Omega A \sim \lambda^2$). 

Based on equation (1), the facility sensitivity grows steeply with aperture diameter ($\propto D^2$) and only weakly with number of telescopes ($\propto \sqrt{N_t}$).  As a dramatic real-world example, the $4\times8$\,m telescopes of the ESO/VLTI UT array has the same background-limited sensitivity for point source detection as an array of $200\times3$\,m telescopes!  Note that this conclusion is subject to some assumptions, as some novel beam combiners \citep[e.g., densified pupil;][]{pedretti2000} can recover the SNR$\propto N_t$ scaling under certain conditions. Furthermore, the angular resolution and capability to do nulling depends on the geometry of the array and so an array of many 3\,m telescopes with long baselines designed for nulling will have superior capabilities in some ways than a $4\times8$\,m array with shorter baselines. {\em  Regardless of details, one sees that it is critical to push for large apertures for sensitivity and that it is not generally cost-effective to try to compensate aperture with more telescopes. } 

\begin{figure}[t]
   \begin{center}
   \includegraphics[width=5.8in]{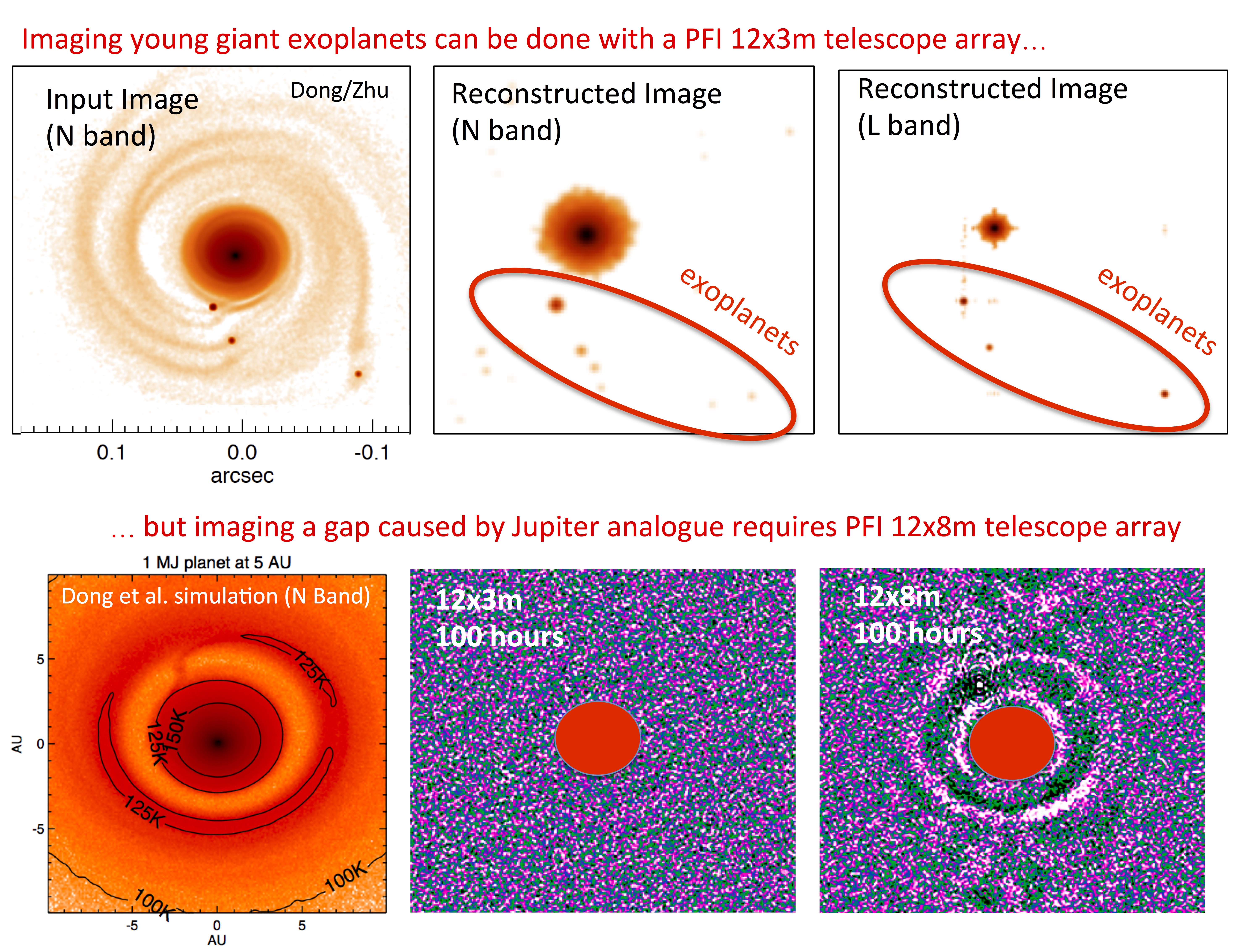}
   \end{center}
   \vspace{-.2in}
   \caption[example] 
   { \label{fig:newsim} 
 \footnotesize The PFI Science and Technical Working Groups have simulated performance for both 12x3\,m and 12x8\,m PFI architectures. (top row) \citet{pfimonnier2016} demonstrated that an equivalent 12x3\,m PFI (in 25 hours) can  detect young giant exoplanets at both L and N bands (``hot start'' models shown here) but cannot see the warm dust in this 4-planet simulation by \citet{dong2015}.  
 (bottom row) \citet{monnier2018pfi} found that a 12x8\,m PFI array would be able to image gap formation in solar analogues found in  nearby star forming regions, a stunning capability bringing ALMA-like performance to the inner solar system where terrestrial planets form.
 }
\end{figure}

One other issue to consider in designing an array is the surface brightness limit for mid-infrared observations.  Since the blackbody function peaks in the mid-IR for $T=300$\,K, we recognize that observing dust much cooler than this in young stellar objects will be fighting the exponential Wien's tail of the Planck function. Quantitatively, emission at $T=150$\,K is $\sim$1\% of the emission at $T=300K$ in the mid-IR.  Indeed, going further to $T=125$\,K is another factor of 7 in flux.  Based on this physics, it is not practical to observe thermal emission of dust much below $T<125$\,K in the mid-infrared at high angular resolution, especially from the ground.  Fortunately, the water iceline occurs at T$\sim$150\,K, opening up observing opportunities for PFI.

Here we motivate the facility characteristics adopted for our PFI reference architectures:
\begin{itemize}
\item 1.2\,km maximum baseline chosen to attain 0.2\,au mid-IR resolution at 140\,pc: a) to resolve a planet-opened gap from a Jupiter at 1\,au, b) to resolve diameters of circumplanetary disks for exo-Jupiter (1\,M$_{J}$@5\,au)
\item $12\times3$\,m diameter array chosen to have $T=150$\,K  surface brightness (3-$\sigma$) in 10,000\,s: 
Sensitivity to dust at $T=150$\,K, the temperature for the water iceline for typical disks
\item $12\times8$\,m diameter array chosen for enhanced surface brightness limit ($T=125$\,K) to see gaps and dust structures for giant planets forming as far out as 5\,au, and an enhanced exoplanet yield.
\item Sufficient point source sensitivity to detect young giant exoplanets for a range of models.
\item Sufficient fringe tracking margin (H band magnitude limit at least 13) to observe solar-type stars in nearby star forming regions.
\end{itemize}

Table\,\ref{pfiref} contains information on two reference facility architectures for PFI, one is a $12\times$3\,m array and the other a $12\times8$\,m array.  Figure~\ref{fig:newsim} shows examples of the science capabilities of the arrays.   The main difference is a factor of 7 in sensitivity that is crucial for imaging warm dust at 5\,au and for a more complete census of giant planets in young disks.  During the PFI Community Meeting held during the 2018 SPIE meeting, a straw poll of those in attendance overwhelmingly favored pursuing the $12\times8$\,m architecture over the $12\times3$\,m version based on the increased science capabilities.

Note that the PFI Science Working Group determined that a mid-latitude site is near-essential for PFI due to the limited number of star-forming regions observable from high-latitude sites, which removes the High Antarctic Plateau from consideration. The PFI Project has identified the Flagstaff (Arizona, USA) Navy Precision Optical Interferometer (NPOI) site and the ALMA site (Chajnantor Plateau, Chile) as locations with sufficient accessible area and existing infrastructure to merit further consideration.

\vspace{0.1in}

\noindent {\bf Technology Drivers:} 

\begin{wrapfigure}[15]{r}{0.5\textwidth}
\vspace{-.8in}
\begin{center}
   \begin{tabular}{c} %
   \includegraphics[height=5cm]{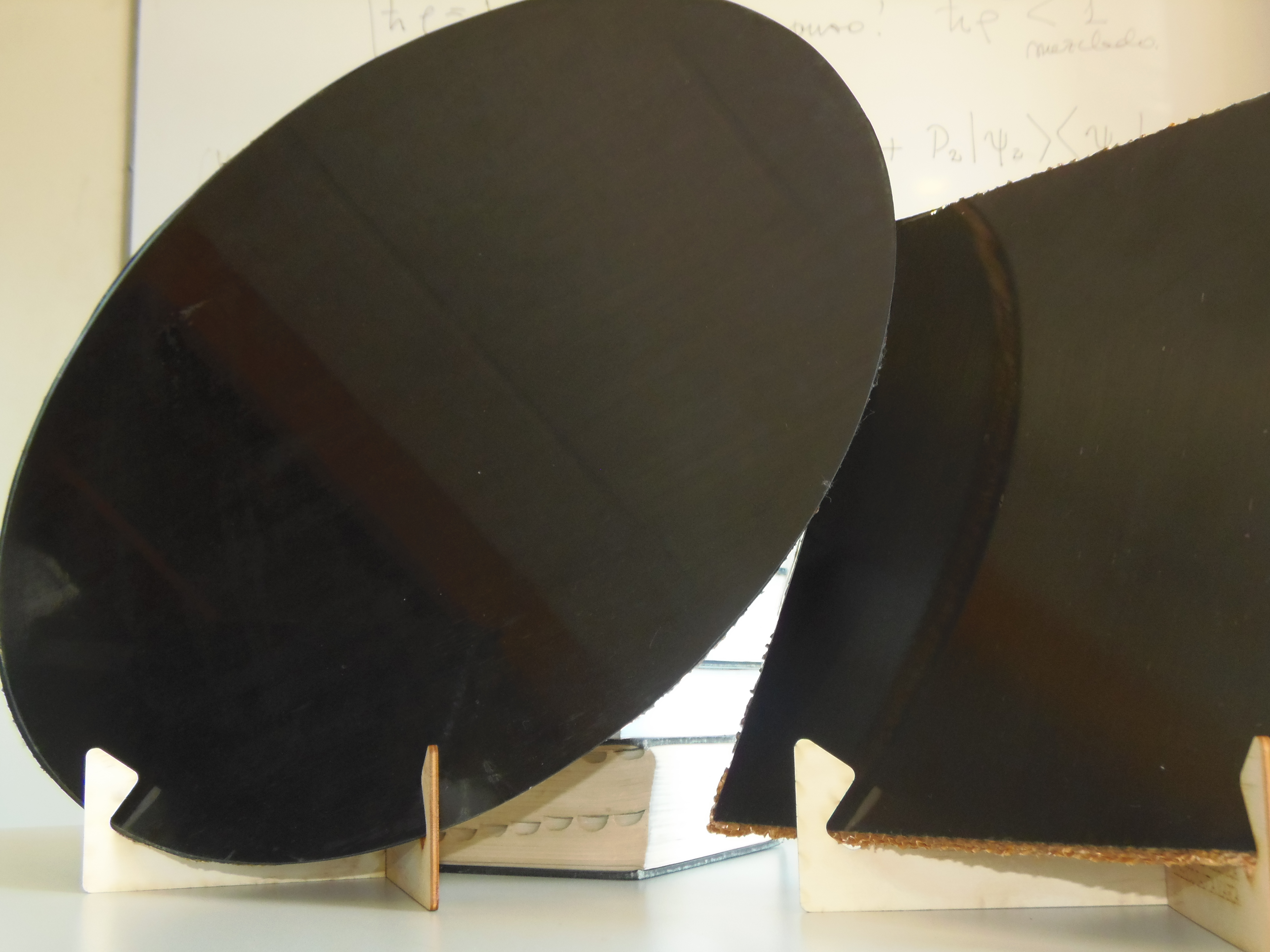}
   \end{tabular}
   \end{center}
   \vspace{-.1in}
   \caption[example] 
   { \label{fig:bayo} 
\footnotesize New technologies to allow mass-production of inexpensive 8\,m class telescopes are essential for the viability of the PFI project long-term.  Here we see 0.5m spherical curvature carbon fiber prototypes made from a stainless steel mandrel under fabrication in Chile \citep{zunigo2018spie}. Credit: Amelia Bayo}
   \end{wrapfigure} 

The PFI Technical Working Group has developed a Technology Roadmap containing technologies identified as strategic and key investments for PFI (see Table~\ref{tab:roadmap}).  We will focus on a few of the most important technology areas next while we refer the interested reader to the recent SPIE summary by \citet{monnier2018spie} for more details.

\vspace{0.1in}

\noindent {\em Technology Driver 1: Inexpensive Telescopes}

The most crucial technical advance needed to make PFI both supremely powerful and affordable is less-expensive large-aperture telescopes. The cost of ``traditional'' telescopes  scale as D$^{2.5}$ \citep{vanbelle2004} -- roughly \$5M for a 2\,m aperture, $\sim$\$40M for a 4\,m aperture -- thus, an array of a dozen or more large telescopes rapidly becomes cost-prohibitive.  A goal of the PFI Technology Program is to attain an order-of-magnitude cost reductions in telescope construction.  

There are three key concepts that drive our optimism that such a breakthrough is possible:
\begin{enumerate}
    \item Advances in lightweight and new ``smart'' materials
    \item Integration of ``active'' and ``adaptive'' optics systems into telescope architecture, enabling imaging within a narrow field-of-view on an intrinsically ``floppy'' structure
    \item Replication savings from mass-production 
\end{enumerate}

Multiple concepts are being explored that take advantage of these principles, including spherical primaries, carbon-fiber reinforced polymers for mirror replication and/or for lightweight supports, slumped glass, and more. Such narrow-field, AO-corrected telescopes would also be useful for upcoming RV surveys of bright stars searching for exo-Earths. Natural allies for this technical development include governments interested in imaging geostationary satellites and telecommunication companies interested in narrow-field, diffraction-limited applications such as laser communication (to/from space or ground stations).  Work is ongoing in Chile \citep[see Figure \ref{fig:bayo} and][]{zunigo2018spie}, USA (Lowell Observatory \& U. Michigan), and Australia (ANU) to seek funding and new partnerships.  For instance, a spherical primary with a highly-aspheric Gregorian secondary produces a Gaussian-like apodization of the pupil with diffraction-limited performance over a small field-of-view \citep{pfimonnier2014,pfiireland2016}. In addition, there could be significant cost reductions even using today's technology through mass production \citep{kingsley2018spie} of proven designs, such as the MMT or Magellan telescopes.

\begin{table}[t]
\caption{PFI Technology Roadmap} 
\label{tab:roadmap}
\begin{center}       
\begin{tabular}{|l|l|}
\hline
Critical Technology &  Considerations \\
\hline
Inexpensive telescopes&  Possible key technologies: \\ 
& \qquad Replicated parabolic or spherical mirrors, \\
& \qquad lightweight structures with AO correction \\
& {\bf Partner with industry, engineers}\\
& {\bf New telescopes for existing arrays (CHARA,NPOI,MROI)}\\
\hline
L/M band Integrated Optics 
& Needed for high precision calibration, \\
combiners & Explore Chalcogenide integrated optics. \\
& {\bf Pilot nulling instruments with VLTI/Hi-5 or MROI} \\
\hline
Wavelength-bootstrapped  
& L band imaging require $10^6$:$1$ dynamic range imaging, \\
fringe tracking 
& ultra-accurate fringe tracking in L based on H-band, \\
 & {\bf VLTI/GRA4MAT mode}\\
& {\bf New ``high sensitivity''  fringe tracker (e.g., VLTI/Heimdallr)}\\
\hline
Mid-IR laser comb heterodyne & Possible ``add-on'' to L/M band \\
 & {\bf Develop combs, detectors, digital processing} \\
\hline
Low-cost operations model  & Demonstrate as part of new 3-telescope Array \\
& {\bf e.g., {\em Gaia}/TESS follow-up interferometer (3x3m)} \\
\hline
Space interferometry & Longer-term future for high sensitivity \\
& Demonstrate formation flying with Cubesats\\
& {\bf Support new formation flying missions using smallSats} \\
\hline
\end{tabular}
\end{center}
\end{table}

\vspace{0.1in}

\noindent {\em Technology Driver 2: Focus on instruments and intermediate facilities}

Over the past two years, a few new proposals for instruments and facilities have been developed that could act as pathfinders for PFI.  Nulling interferometry is needed for detection of exoplanets with PFI at short wavelengths and the Hi-5 project is a science-driven international initiative to develop a new VLTI instrument optimized for high dynamic range observations \citep{Defrere:2018}.  Reaching 1:10,000 or higher dynamic range in the thermal near-infrared (L and M bands) would be extremely valuable to directly detect forming and young giant exoplanets.  Nulling interferometry on the ground also requires advances in highly-accurate fringe tracking and VLTI could demonstrate this with the proposed Heimdallr instrument (PI: Ireland).
While no comparable instruments have been proposed yet for US facilities, an L/M band nulling instrument could potentially be interesting at MROI using 4 telescopes before the full 10-telescope array is available.

MATISSE is a L/M/N band 4-telescope combiner for VLTI that was just commissioned \citep{lopez2018spie}.  MATISSE should be able to be used with GRAVITY in fringe-tracking mode (GRA4MAT) and reach very high sensitivity, although with  much lower angular resolution than PFI.
While VLTI has relatively short baselines (120\,m instead of 1200\,m needed for PFI), the sensitivity of 4x8\,m telescopes is unrivaled  -- this ``mini-PFI" could be an amazing testbed for future PFI technology developments and science instruments. The PFI Project strongly advocates developing the VLTI-UT interferometric capabilities during the 2020s, ideally with involvement from USA astronomers.

\vspace{0.2in}
\noindent {\em Technology Driver 3: Gaia/TESS follow-up with a new $3\times 3$\,m facility with 1\,km baselines}

\begin{wrapfigure}[20]{r}{0.5\textwidth}
\vspace{-.2in}
  \includegraphics[width=.5\textwidth]{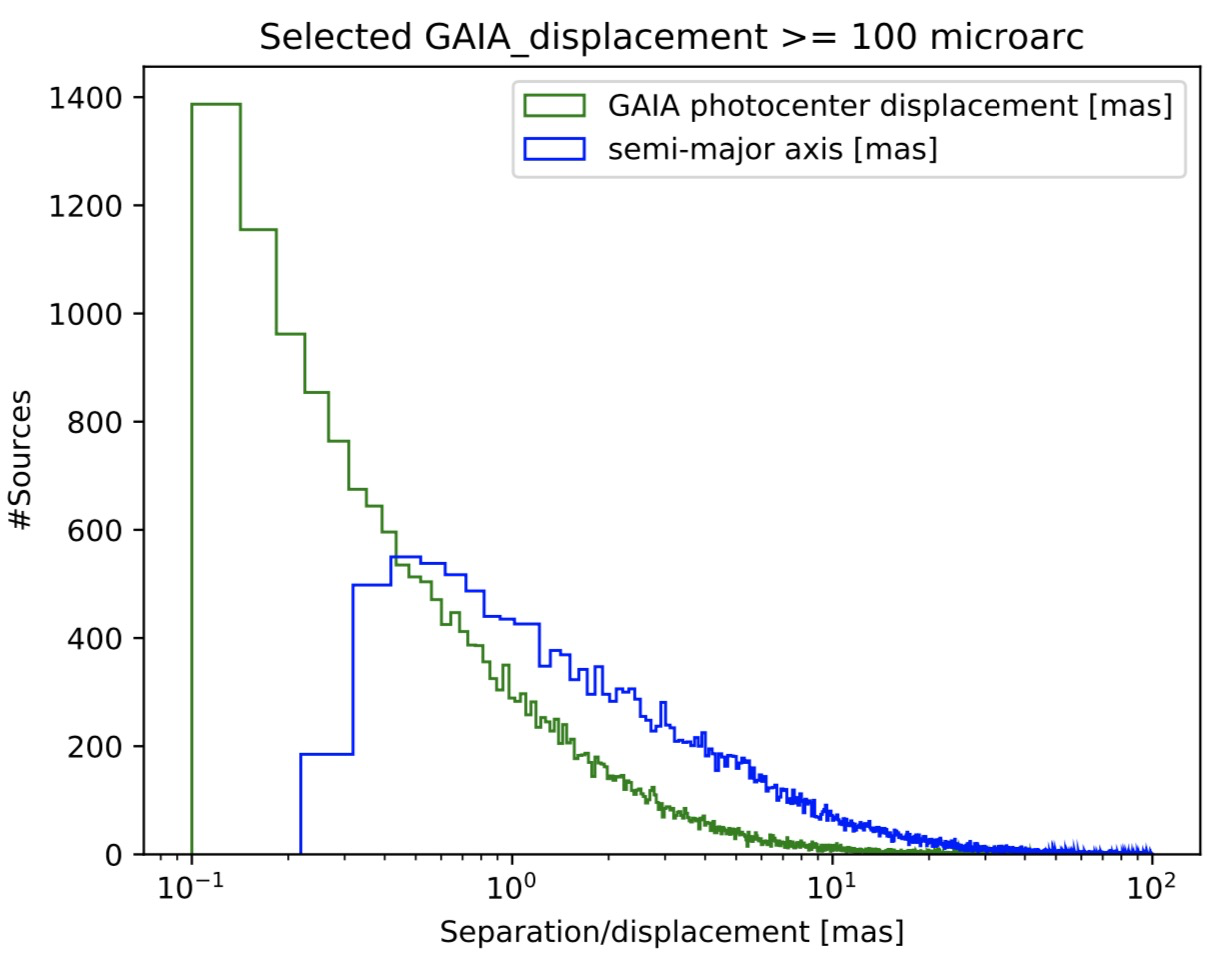}
\caption{\footnotesize Using a Galactic structure simulation, we can estimate the binary properties detected by {\em Gaia}.  From this, we can estimate the angular resolution needed for a ground-based interferometer to resolve these systems -- thus providing absolute masses for both components.  We see there are {\em tens of thousands} of objects that can by resolved even with current facilities (Ireland et al., in prep). }
\label{fig:2}       %
\end{wrapfigure}

While exoplanet science and thermal dust imaging of cool dust requires large aperture telescopes, a small array of $3\times3$\,m telescopes with $\sim$1\,km baselines could test key PFI technologies such as cheap telescopes, long delay lines, and low-budget operations model while achieving impressive science goals. In particular, {\em Gaia} and TESS will identify millions of nearby binary stars and red giants that need additional data. By measuring the separation of {\em Gaia} binaries with photocenter orbits, a $3\times3$\,m interferometer would measure thousands of new masses for a wide range of stars in the solar neighborhood. Furthermore, diameters of TESS red giants with asteroseismic radii from precision photometry will allow independent calibration of this key technique.

To illustrate this, we have estimated source counts using the Casagrande simulation (Ireland et al., in prep.), applying the following selection criteria: (1) period $<$ 10\,years,
(2) $R<15$ mag (apparent magnitude),
(3) {\em Gaia} photocenter displacement $>100$\,microarcsec (i.e. SNR$\approx10$ detection of photocenter displacement). In total, there are 27,464 sources that match these criteria and 26,332 have semi-major axis $>$ 0.5\,mas (i.e. in range of current CHARA and VLTI facilities).  A new VLTI instrument, BIFROST, operating at 1\,$\mu$m is being proposed (Kraus, Ireland) to observe the wider {\em Gaia} binaries.  A new facility with kilometric baselines will resolve additional binaries at smaller separations and permit robust effective temperature calibrations across the HR diagram, including stars with very low metallicity.

\newpage
\noindent {\em Technology Driver 4: Space Interferometry}

Lastly, space interferometry has the potential to extend ``PFI science'' to the low-mass (possibly Earth-mass) planetary mass regime. In addition, through cooled apertures, ``Space-PFI'' has the potential to fully explore the 5--20\,au regime where disk gaps remain resolution-limited for ALMA and sensitivity-limited for a ground-based PFI.  The PFI project strongly supports taking some steps in the next few years to explore Cubesat formation flying missions, to launch a low-cost space interferometer demonstration, and to participate in NASA/ESA planning processes to explore the massive advantages of developing interferometric space capabilities for missions from far-infrared to x-rays.  {\em There will be other ASTRO2020 Technical White Papers on this topic ``A Realistic Roadmap to Formation Flying Space Interferometer" (Lead: Monnier) and ``A Long-Term Vision for Space-Based Interferometry'' (Lead: Rinehart).}

\vspace{0.1in}

\noindent {\bf Organization, Partnerships, and Current Status:} 

The Planet Formation Imager Project is international with various collaborations already formed to work on PFI technology.  In the context of the USA ASTRO2020 Decadal Survey, we emphasize that US-led projects can take advantage of the longest interferometer baselines in the world at CHARA,  NPOI, and MROI, and these facilities have pioneered many new technologies, from new telescope designs to new beam combiners to new delay line designs.  

\vspace{0.1in}

\noindent {\bf Schedule:} 

As we are proposing technology development, we do not advocate a detailed timeline for construction of the Planet Formation Imager (PFI) as a full facility.
The most critical advance required for PFI is less expensive $5+$m apertures.  As a goal for this decade, we would like to see inexpensive 3\,m-class telescopes ($<\$2M$ each) built from technologies traceable up to 8\,m class.  This kind of order-of-magnitude reduction in cost would make PFI a competitive project for ASTRO2030.

A medium-scale  project this decade to build a 3x3\,m telescope facility with kilometric baselines would tackle {\em Gaia}/TESS follow-up science while simultaneously demonstrating key aspects of PFI -- inexpensive telescopes, kilometric baseline beam combination, and low-cost operations. 

\vspace{0.1in}

\noindent {\bf Cost Estimates:} 

Many technologies for PFI can be developed through normal NSF programs.  However, developing new telescope technologies will require close academic-industry ties, requiring larger than usual budgets involving skilled engineers. \$10M would allow serious prototyping of one or two different design concepts at the 1\,m class with a goal of selecting a technology concept for a 3\,m diameter demonstrator at the end of the decade.   We note again that other areas of astronomy would benefit from inexpensive 3\,m class diffraction-limited, narrow-field telescopes, such as next-generation ambitious exoplanet RV surveys.

The cost of a new facility with 3x3\,m telescopes with kilometric baselines depends on many factors.  Adding telescopes to existing facilities is one possibility though geographical site limitations and limited delay line lengths pose serious issues for all current facilities.  Our cost model \citep{ireland2016spie} for a new facility includes: 3 delay lines (\$2.5M), 3 AO systems (\$3.4M), beam combiners (\$6M), other infrastructure (\$1M), and 3x3\,m telescopes (cost from \$6M to \$26M, depending on cost model). With 10\% management and 20\% contingency, we arrive at a facility cost estimate ranging from \$25M to \$50M, with up to half the cost in telescopes alone if no breakthroughs in telescope costs can be achieved this decade.  Such a project would benefit from international partnerships to share cost.

\pagebreak
{\small
\bibliography{ASTRO2020}

\begin{thebibliography}{}

\bibitem[{Besser} et~al., 2018]{besser2018spie}
{Besser}, F.~E., {Ramos}, N., {Rates}, A., {Ortega}, N., {Sepulveda}, S.,
  {Parvex}, T., {Pi{\~n}a}, M., {Pollarolo}, C., {Jara}, R.~E., and {Espinoza},
  K.~R. (2018).
\newblock {Fiber-based infrared heterodyne technology for the PFI: development
  of a prototype test system}.
\newblock In {\em \procspie}, volume 10701 of {\em Society of Photo-Optical
  Instrumentation Engineers (SPIE) Conference Series}, page 107012L.

\bibitem[{Besser} et~al., 2016]{pfibesser2016}
{Besser}, F.~E., {Rates}, A., {Ortega}, N., {Pina}, M.~I., {Pollarolo}, C.,
  {Jofre}, M., {Ya{\~n}ez}, C., {Lasen}, M., {Ramos}, N., and {Michael}, E.~A.
  (2016).
\newblock {Fiber-based heterodyne infrared interferometry: an instrumentation
  study platform on the way to the proposed Infrared Planet Formation Imager}.
\newblock In {\em Optical and Infrared Interferometry and Imaging V}, volume
  9907 of {\em \procspie}, page 99072L.

\bibitem[{Defr{\`e}re} et~al., 2018a]{Defrere:2018}
{Defr{\`e}re}, D., {Absil}, O., {Berger}, J.-P., {Boulet}, T., {Danchi}, W.~C.,
  {Ertel}, S., {Gallenne}, A., {H{\'e}nault}, F., {Hinz}, P., {Huby}, E.,
  {Ireland}, M., {Kraus}, S., {Labadie}, L., {Le Bouquin}, J.-B., {Martin}, G.,
  {Matter}, A., {M{\'e}rand}, A., {Mennesson}, B., {Minardi}, S., {Monnier},
  J., {Norris}, B., {Orban de Xivry}, G., {Pedretti}, E., {Pott}, J.-U.,
  {Reggiani}, M., {Serabyn}, E., {Surdej}, J., {Tristram}, K.~R.~W., and
  {Woillez}, J. (2018a).
\newblock {The path towards high-contrast imaging with the VLTI: the Hi-5
  project}.
\newblock {\em ArXiv e-prints}.

\bibitem[{Defr{\`e}re} et~al., 2018b]{defrere2018spie}
{Defr{\`e}re}, D., {Ireland}, M., {Absil}, O., {Berger}, J.~P., {Danchi},
  W.~C., {Ertel}, S., {Gallenne}, A., {H{\'e}nault}, F., {Hinz}, P., and
  {Huby}, E. (2018b).
\newblock {Hi-5: a potential high-contrast thermal near-infrared imager for the
  VLTI}.
\newblock In {\em \procspie}, volume 10701 of {\em Society of Photo-Optical
  Instrumentation Engineers (SPIE) Conference Series}, page 107010U.

\bibitem[{Dong} et~al., 2015]{dong2015}
{Dong}, R., {Zhu}, Z., and {Whitney}, B. (2015).
\newblock {Observational Signatures of Planets in Protoplanetary Disks I. Gaps
  Opened by Single and Multiple Young Planets in Disks}.
\newblock {\em \apj}, 809:93.

\bibitem[{Ireland} et~al., 2018]{ireland2018spie}
{Ireland}, M.~J., {Defr{\`e}re}, D., {Martinache}, F., {Monnier}, J.~D.,
  {Norris}, B., {Tuthill}, P., and {Woillez}, J. (2018).
\newblock {Image-plane fringe tracker for adaptive-optics assisted long
  baseline interferometry}.
\newblock In {\em \procspie}, volume 10701 of {\em Society of Photo-Optical
  Instrumentation Engineers (SPIE) Conference Series}, page 1070111.

\bibitem[{Ireland} and {Monnier}, 2014]{pfiireland2014}
{Ireland}, M.~J. and {Monnier}, J.~D. (2014).
\newblock {A dispersed heterodyne design for the planet formation imager}.
\newblock In {\em Optical and Infrared Interferometry IV}, volume 9146 of {\em
  \procspie}, page 914612.

\bibitem[{Ireland} et~al., 2016a]{ireland2016spie}
{Ireland}, M.~J., {Monnier}, J.~D., {Kraus}, S., {Isella}, A., {Minardi}, S.,
  {Petrov}, R., {ten Brummelaar}, T., {Young}, J., {Vasisht}, G., and
  {Mozurkewich}, D. (2016a).
\newblock {Status of the Planet Formation Imager (PFI) concept}.
\newblock In {\em \procspie}, volume 9907 of {\em Society of Photo-Optical
  Instrumentation Engineers (SPIE) Conference Series}, page 99071L.

\bibitem[{Ireland} et~al., 2016b]{pfiireland2016}
{Ireland}, M.~J., {Monnier}, J.~D., {Kraus}, S., {Isella}, A., {Minardi}, S.,
  {Petrov}, R., {ten Brummelaar}, T., {Young}, J., {Vasisht}, G.,
  {Mozurkewich}, D., {Rinehart}, S., {Michael}, E.~A., {van Belle}, G., and
  {Woillez}, J. (2016b).
\newblock {Status of the Planet Formation Imager (PFI) concept}.
\newblock In {\em Optical and Infrared Interferometry and Imaging V}, volume
  9907 of {\em \procspie}, page 99071L.

\bibitem[{Kingsley} et~al., 2018]{kingsley2018spie}
{Kingsley}, J.~S., {Angel}, R., {Davison}, W., {Neff}, D., {Teran}, J.,
  {Assenmacher}, B., {Peyton}, K., {Martin}, H.~M., {Oh}, C., and {Kim}, D.
  (2018).
\newblock {An inexpensive turnkey 6.5m observatory with customizing options}.
\newblock In {\em \procspie}, volume 10700 of {\em Society of Photo-Optical
  Instrumentation Engineers (SPIE) Conference Series}, page 107004H.

\bibitem[{Kraus} et~al., 2014]{pfikraus2014}
{Kraus}, S., {Monnier}, J., {Harries}, T., {Dong}, R., {Bate}, M., {Whitney},
  B., {Zhu}, Z., {Buscher}, D., {Berger}, J.-P., {Haniff}, C., {Ireland}, M.,
  {Labadie}, L., {Lacour}, S., {Petrov}, R., {Ridgway}, S., {Surdej}, J., {ten
  Brummelaar}, T., {Tuthill}, P., and {van Belle}, G. (2014).
\newblock {The science case for the Planet Formation Imager (PFI)}.
\newblock In {\em Optical and Infrared Interferometry IV}, volume 9146 of {\em
  \procspie}, page 914611.

\bibitem[{Kraus} et~al., 2016]{pfikraus2016}
{Kraus}, S., {Monnier}, J.~D., {Ireland}, M.~J., {Duch{\^e}ne}, G.,
  {Espaillat}, C., {H{\"o}nig}, S., {Juhasz}, A., {Mordasini}, C., {Olofsson},
  J., {Paladini}, C., {Stassun}, K., {Turner}, N., {Vasisht}, G., {Harries},
  T.~J., {Bate}, M.~R., {Gonzalez}, J.-F., {Matter}, A., {Zhu}, Z., {Panic},
  O., {Regaly}, Z., {Morbidelli}, A., {Meru}, F., {Wolf}, S., {Ilee}, J.,
  {Berger}, J.-P., {Zhao}, M., {Kral}, Q., {Morlok}, A., {Bonsor}, A.,
  {Ciardi}, D., {Kane}, S.~R., {Kratter}, K., {Laughlin}, G., {Pepper}, J.,
  {Raymond}, S., {Labadie}, L., {Nelson}, R.~P., {Weigelt}, G., {ten
  Brummelaar}, T., {Pierens}, A., {Oudmaijer}, R., {Kley}, W., {Pope}, B.,
  {Jensen}, E.~L.~N., {Bayo}, A., {Smith}, M., {Boyajian}, T.,
  {Quiroga-Nu{\~n}ez}, L.~H., {Millan-Gabet}, R., {Chiavassa}, A., {Gallenne},
  A., {Reynolds}, M., {de Wit}, W.-J., {Wittkowski}, M., {Millour}, F.,
  {Gandhi}, P., {Ramos Almeida}, C., {Alonso Herrero}, A., {Packham}, C.,
  {Kishimoto}, M., {Tristram}, K.~R.~W., {Pott}, J.-U., {Surdej}, J.,
  {Buscher}, D., {Haniff}, C., {Lacour}, S., {Petrov}, R., {Ridgway}, S.,
  {Tuthill}, P., {van Belle}, G., {Armitage}, P., {Baruteau}, C., {Benisty},
  M., {Bitsch}, B., {Paardekooper}, S.-J., {Pinte}, C., {Masset}, F., and
  {Rosotti}, G. (2016).
\newblock {Planet Formation Imager (PFI): science vision and key requirements}.
\newblock In {\em Optical and Infrared Interferometry and Imaging V}, volume
  9907 of {\em \procspie}, page 99071K.

\bibitem[{Lopez} et~al., 2018]{lopez2018spie}
{Lopez}, B., {Lagarde}, S., {Matter}, A., {Agocs}, T., {Allouche}, F.,
  {Antonelli}, P., {Augereau}, J.~C., {Bailet}, C., {Berio}, P., and
  {Bettonvil}, F. (2018).
\newblock {The installation and ongoing commissioning of the MATISSE
  mid-infrared interferometer at the ESO Very Large Telescope Observatory}.
\newblock In {\em \procspie}, volume 10701 of {\em Society of Photo-Optical
  Instrumentation Engineers (SPIE) Conference Series}, page 107010Z.

\bibitem[{Michael} and {Besser}, 2018]{michael2018spie}
{Michael}, E.~A. and {Besser}, F.~E. (2018).
\newblock {Fiber-based infrared heterodyne technology for the PFI: on the
  possibility of breaking the noise temperature quantum limit with
  cross-correlation}.
\newblock In {\em \procspie}, volume 10701 of {\em Society of Photo-Optical
  Instrumentation Engineers (SPIE) Conference Series}, page 107011X.

\bibitem[{Minardi} et~al., 2016]{pfiminardi2016}
{Minardi}, S., {Lacour}, S., {Berger}, J.-P., {Labadie}, L., {Thomson}, R.~R.,
  {Haniff}, C., and {Ireland}, M. (2016).
\newblock {Beam combination schemes and technologies for the Planet Formation
  Imager}.
\newblock In {\em Optical and Infrared Interferometry and Imaging V}, volume
  9907 of {\em \procspie}, page 99071N.

\bibitem[{Monnier} et~al., 2018a]{monnier2018spie}
{Monnier}, J.~D., {Ireland}, M., {Kraus}, S., {Alonso-Herrero}, A., {Bonsor},
  A., {Baron}, F., {Bayo}, A., {Berger}, J.-P., {Boyajian}, T., and
  {Chiavassa}, A. (2018a).
\newblock {Planet formation imager: project update}.
\newblock In {\em \procspie}, volume 10701 of {\em Society of Photo-Optical
  Instrumentation Engineers (SPIE) Conference Series}, page 1070118.

\bibitem[{Monnier} et~al., 2016]{pfimonnier2016}
{Monnier}, J.~D., {Ireland}, M.~J., {Kraus}, S., {Baron}, F., {Creech-Eakman},
  M., {Dong}, R., {Isella}, A., {Merand}, A., {Michael}, E., {Minardi}, S.,
  {Mozurkewich}, D., {Petrov}, R., {Rinehart}, S., {ten Brummelaar}, T.,
  {Vasisht}, G., {Wishnow}, E., {Young}, J., and {Zhu}, Z. (2016).
\newblock {Architecture design study and technology road map for the Planet
  Formation Imager (PFI)}.
\newblock In {\em Optical and Infrared Interferometry and Imaging V}, volume
  9907 of {\em \procspie}, page 99071O.

\bibitem[{Monnier} et~al., 2014]{pfimonnier2014}
{Monnier}, J.~D., {Kraus}, S., {Buscher}, D., {Berger}, J.-P., {Haniff}, C.,
  {Ireland}, M., {Labadie}, L., {Lacour}, S., {Le Coroller}, H., {Petrov},
  R.~G., {Pott}, J.-U., {Ridgway}, S., {Surdej}, J., {ten Brummelaar}, T.,
  {Tuthill}, P., and {van Belle}, G. (2014).
\newblock {Planet formation imager (PFI): introduction and technical
  considerations}.
\newblock In {\em Optical and Infrared Interferometry IV}, volume 9146 of {\em
  \procspie}, page 914610.

\bibitem[{Monnier} et~al., 2018b]{monnier2018pfi}
{Monnier}, J.~D., {Kraus}, S., {Ireland}, M.~J., {Baron}, F., {Bayo}, A.,
  {Berger}, J.-P., {Creech-Eakman}, M., {Dong}, R., {Duch{\^e}ne}, G., and
  {Espaillat}, C. (2018b).
\newblock {The planet formation imager}.
\newblock {\em Experimental Astronomy}, 46(3):517--529.

\bibitem[{Mozurkewich} et~al., 2016]{pfimozurkewich2016}
{Mozurkewich}, D., {Young}, J., and {Ireland}, M. (2016).
\newblock {Practical Beam Transport for the Planet Formation Imager (PFI)}.
\newblock {\em ArXiv e-prints}.

\bibitem[{Pedretti} et~al., 2018]{pedretti2018spie}
{Pedretti}, E., {Diener}, R., {Shankar Nayak}, A., {Tepper}, J., {Labadie}, L.,
  {Pertsch}, T., {Nolte}, S., and {Minardi}, S. (2018).
\newblock {Beam combination schemes and technologies for the Planet Formation
  Imager}.
\newblock In {\em \procspie}, volume 10701 of {\em Society of Photo-Optical
  Instrumentation Engineers (SPIE) Conference Series}, page 107012O.

\bibitem[{Pedretti} et~al., 2000]{pedretti2000}
{Pedretti}, E., {Labeyrie}, A., {Arnold}, L., {Thureau}, N., {Lardiere}, O.,
  {Boccaletti}, A., and {Riaud}, P. (2000).
\newblock {First images on the sky from a hyper telescope}.
\newblock {\em \aaps}, 147:285--290.

\bibitem[{Petrov} et~al., 2016]{pfipetrov2016}
{Petrov}, R.~G., {Boskri}, A., {Elhalkouj}, T., {Monnier}, J., {Ireland}, M.,
  and {Kraus}, S. (2016).
\newblock {Co-phasing the planet formation imager}.
\newblock In {\em Optical and Infrared Interferometry and Imaging V}, volume
  9907 of {\em \procspie}, page 99073W.

\bibitem[{Quanz} et~al., 2018]{quanz2018spie}
{Quanz}, S.~P., {Kammerer}, J., {Defr{\`e}re}, D., {Absil}, O., {Glauser},
  A.~M., and {Kitzmann}, D. (2018).
\newblock {Exoplanet science with a space-based mid-infrared nulling
  interferometer}.
\newblock In {\em \procspie}, volume 10701 of {\em Society of Photo-Optical
  Instrumentation Engineers (SPIE) Conference Series}, page 107011I.

\bibitem[{Raymond} et~al., 2006]{raymond2006}
{Raymond}, S.~N., {Quinn}, T., and {Lunine}, J.~I. (2006).
\newblock {High-resolution simulations of the final assembly of Earth-like
  planets I. Terrestrial accretion and dynamics}.
\newblock {\em Icarus}, 183:265--282.

\bibitem[{Tristram} and {H{\"o}nig}, 2018]{tristam2018spie}
{Tristram}, K. R.~W. and {H{\"o}nig}, S.~F. (2018).
\newblock {The success of extragalactic infrared interferometry: from what we
  have learned to what to expect}.
\newblock In {\em \procspie}, volume 10701 of {\em Society of Photo-Optical
  Instrumentation Engineers (SPIE) Conference Series}, page 107011V.

\bibitem[{van Belle} et~al., 2004]{vanbelle2004}
{van Belle}, G.~T., {Meinel}, A.~B., and {Meinel}, M.~P. (2004).
\newblock {The scaling relationship between telescope cost and aperture size
  for very large telescopes}.
\newblock In {Oschmann}, Jr., J.~M., editor, {\em Ground-based Telescopes},
  volume 5489 of {\em \procspie}, pages 563--570.

\bibitem[{Z{\'u}{\~n}iga-Fern{\'a}ndez} et~al., 2018]{zunigo2018spie}
{Z{\'u}{\~n}iga-Fern{\'a}ndez}, S., {Bayo}, A., {Olofsson}, J., {Pedrero}, L.,
  {Lobos}, C., {Rozas}, E., {Soto}, N., {Schreiber}, M., {Esc{\'a}rate}, P.,
  and {Romero}, C. (2018).
\newblock {NPF: mirror development in Chile}.
\newblock In {\em \procspie}, volume 10700 of {\em Society of Photo-Optical
  Instrumentation Engineers (SPIE) Conference Series}, page 107003X.

\end{thebibliography}
\bibliographystyle{apalike}
}

\end{document}